\documentclass[superscriptaddress,citeautoscript,floatfix,longbibliography,bibnotes,nofootinbib]{revtex4-2}
\usepackage{times,bm,amsmath,amsfonts,amsthm,amssymb,amsbsy,braket,graphicx}
\usepackage[utf8]{inputenc}
\usepackage[T1]{fontenc}
\usepackage{hyperref}
\usepackage[capitalize]{cleveref}
\usepackage{microtype}
\newcommand{\ii}{\mathrm{i}}
\newcommand{\ee}{\mathrm{e}}
\newcommand{\dd}{\mathrm{d}}
\renewcommand{\Re}{\mathrm{Re}}
\renewcommand{\Im}{\mathrm{Im}}
\DeclareMathOperator{\sgn}{sgn}
\DeclareMathOperator{\const}{const}
\DeclareMathOperator{\sech}{sech}
\DeclareMathOperator{\erf}{erf}
\DeclareMathOperator{\erfi}{erfi}
\hypersetup{colorlinks=true,urlcolor=blue,linkcolor=blue,citecolor=blue}

\begin{document}
\title{
Zero energy modes with Gaussian, exponential, or polynomial decay:
\\
Exact solutions in hermitian and nonhermitian regimes
}
\author{Pasquale Marra}
\email{pasquale.marra@keio.jp}
\affiliation{Graduate School of Mathematical Sciences, The University of Tokyo, 3-8-1 Komaba, Meguro, Tokyo, 153-8914, Japan}
\affiliation{Department of Physics, and Research and Education Center for Natural Sciences, Keio University, 4-1-1 Hiyoshi, Yokohama, Kanagawa, 223-8521, Japan}
\author{Angela Nigro}
\email{anigro@unisa.it}
\affiliation{
Dipartimento di Fisica ``E. R. Caianiello'', Università degli Studi di Salerno, 84084 Fisciano (Salerno), Italy}

\begin{abstract}
Topological zero modes in topological insulators or superconductors are exponentially localized at the phase transition between a topologically trivial and nontrivial phase. These modes are solutions of a Jackiw-Rebbi equation modified with an additional term which is quadratic in the momentum. 
Moreover, localized fermionic modes can also be induced by harmonic potentials in superfluids and superconductors or in atomic nuclei.
Here, by using inverse methods, we consider in the same framework exponentially-localized zero modes, as well as Gaussian modes induced by harmonic potentials (with superexponential decay) and polynomially decaying modes (with subexponential decay), and derive the explicit and analytical form of the modified Jackiw-Rebbi equation (and of the Schrödinger equation) which admits these modes as solutions.
We find that the asymptotic behavior of the mass term is crucial in determining the decay properties of the modes.
Furthermore, these considerations naturally extend to the nonhermitian regime.
These findings allow us to classify and understand topological and nontopological boundary modes in topological insulators and superconductors.  
\end{abstract}
\maketitle

\section{Introduction}

Topological edge modes are exponentially localized modes that appear at the domain wall between topologically non-equivalent phases, e.g., at the edges of a topological insulator or superconductor~\cite{schnyder_classification_2009,hasan_colloquium:_2010,qi_topological_2011,shen_topological_2011,wehling_dirac_2014,shen_topological_2017}, e.g., quantum Hall insulators~\cite{thouless_quantized_1982}, or superfluids~\cite{volovik_fermion_1999,read_non-abelian_2009}.
In systems with charge conjugation symmetry (particle-hole symmetry in condensed matter language), these modes are pinned to zero energy or have energy approaching zero exponentially in the limit of infinite system sizes.
Zero energy modes appear in topological superfluids such as superfluid helium-3~\cite{leggett_a-theoretical_1975,volovik_the-universe_2009}, or as boundary modes~\cite{shen_topological_2017} of topological superconductors~\cite{qi_topological_2011,hasan_colloquium:_2010}, in particular in  one-dimensional topological superconductors~\cite{read_paired_2000,kitaev_unpaired_2001,fu_superconducting_2008,sato_non-abelian_2009,sato_non-abelian_2010,alicea_majorana_2010,lutchyn_majorana_2010,oreg_helical_2010} (see Refs.~\onlinecite{alicea_new-directions_2012,leijnse_introduction_2012,tanaka_symmetry_2012,stanescu_majorana_2013,beenakker_search_2013,elliott_colloquium:_2015,das-sarma_majorana_2015,sato_majorana_2016,sato_topological_2017,aguado_majorana_2017,laubscher_majorana_2021,marra_majorana_2022,masaki_non-abelian_2024,tanaka_theory_2024} for reviews).
These modes are typically solutions of a modified Jackiw-Rebbi equation, which generally describes the Hamiltonian of the low-energy sector of topological insulators and superconductors:
This is the usual Jackiw-Rebbi equation~\cite{jackiw_solitons_1976} with the addition of a quadratic term in the momentum, and where the mass term and the Dirac velocity are spatially dependent.
The sign of the mass term and of the Dirac velocity determines the topological invariant~\cite{shen_topological_2011,shen_topological_2017}.
The zero modes are localized domain walls that separate regions with opposite masses, i.e., different topological invariants, as a consequence of the bulk-boundary correspondence~\cite{hatsugai_chern_1993,ryu_topological_2002,teo_topological_2010}.
The wavefunctions of these modes are known analytically in the case of a sharp domain wall, i.e., when the mass term is a Heaviside step function~\cite{shen_topological_2011,shen_topological_2017}.
A similar example of exponentially localized modes are skin modes in nonhermitian systems, which can also be interpreted in the framework of topology and bulk-boundary correspondence~\cite{yuto-ashida_non-hermitian_2020,okuma_non-hermitian_2023}.
Recently, we derived the analytical wavefunctions in the case of a smooth domain wall which is exponentially confined with a finite localization length $w$~\cite{marra_topological_2024}.
In all the cases considered, the zero modes are characterized by a well-definite pseudospin $s=\pm1$, which determines the symmetry or antisymmetry of the wavefunction with respect to particle and antiparticle sectors (particle and hole in condensed matter).
We also unveiled a duality between zero-energy eigenmodes of the modified Jackiw-Rebbi equation and the energy eigenmodes of the Schrödinger equation~\cite{marra_topological_2024}.
 
However, exponentially localized modes are not the only possibility:
In principle, the zero modes of the modified Jackiw-Rebbi equation (describing topological insulators and superconductors) can have faster-than-exponential (superexponential) or slower-than-exponential (subexponential) localization properties.
In particular, localized Gaussian modes are induced by harmonic (quadratic) scalar potentials.
Harmonic potentials have been studied in condensed matter, in particular in the case of superfluids of cold atoms~\cite{zhou_revealing_2013,kheirkhah_majorana_2020} or second-order topological superconductors~\cite{kheirkhah_majorana_2020} confined by a harmonic trap, as well as in nuclear and particle physics:
In nuclear shell models, indeed, harmonic scalar (or vector) potentials describe nuclear mean-field theories of nucleons interacting with mesons~\cite{wen-chao_bound_2002,ti-sheng_pseudospin_2003,alberto_tensor_2005,akcay_dirac_2009,akcay_exact_2009}.
Another possibility is zero modes with periodic wavefunctions.
Similar considerations can be applied to the Schrödinger equation with spatially dependent potential, which can exhibit, in principle, energy eigenmodes with faster-than-exponential, exponential, or slower-than-exponential localization properties.

Here, we investigate zero energy modes of the modified Jackiw-Rebbi equation and the eigenmodes of the Schrödinger equation considering different localization properties (faster-than-exponential, exponential, or slower-than-exponential) in the hermitian and nonhermitian regimes and find general relations between these properties and the asymptotic behavior of the fields at large distances, by solving the equations analytically via inverse methods.
These relations are consistent with, but are not a simple consequence of the topological bulk-boundary correspondence.
These findings are relevant for classifying the zero energy modes in topological insulators and superconductors, e.g., Majorana zero modes, and particularly distinguish between modes which are topologically protected by a topological phase transition, which are exponentially localized~\cite{read_non-abelian_2009,kitaev_unpaired_2001,alicea_new-directions_2012,leijnse_introduction_2012,das-sarma_majorana_2015,sato_majorana_2016,aguado_majorana_2017,laubscher_majorana_2021,marra_majorana_2022,tanaka_theory_2024}, 
and other zero modes arising from disorder and inhomogeneities, which lack exponential localization~\cite{kells_near-zero-energy_2012,stanescu_disentangling_2013,liu_andreev_2017,hell_distinguishing_2018,liu_distinguishing_2018,penaranda_quantifying_2018,avila_non-hermitian_2019,vuik_reproducing_2019,yavilberg_differentiating_2019,prada_from_2020,pan_physical_2020,pan_crossover_2021,tian_distinguishing_2021,cayao_distinguishing_2021,liu_majorana_2021,marra_majorana/andreev_2022}, such as trivial Andreev bound states~\cite{kells_near-zero-energy_2012,stanescu_disentangling_2013,liu_andreev_2017,hell_distinguishing_2018,liu_distinguishing_2018,penaranda_quantifying_2018,avila_non-hermitian_2019,vuik_reproducing_2019,yavilberg_differentiating_2019,prada_from_2020,pan_physical_2020,pan_crossover_2021,tian_distinguishing_2021,cayao_distinguishing_2021,liu_majorana_2021,marra_majorana/andreev_2022}.

\section{Summary of the Results}

In order to investigate the multitude of many and various properties of the modified Jackiw-Rebbi equation and the eigenmodes of the Schrödinger equation, we use an inverse approach, which we describe here in general terms.
We start by considering examples of zero mode wavefunctions $\varphi_1(x)$ with faster-than-exponential, exponential, or slower-than-exponential decay, specifically, Gaussian modes, exponentially-localized modes, modes with polynomial decay, and periodic modes.
We thus derive the spatial dependence of the mass $m(x)$ and of the Dirac velocity $v(x)$, such that the ensuing modified Jackiw-Rebbi equation has the mode $\varphi_1(x)$ as a solution.
We then derive the spatial dependence of the potential $V(x)$, such that the ensuing Schrödinger equation has these modes as solutions.
Furthermore, we obtain a second linearly independent eigenmode satisfying the same modified Jackiw-Rebbi equation and with the same pseudospin using the reduction of order technique.
Moreover, using explicitly the duality between zero eigenmodes of the modified Jackiw-Rebbi equation and the energy eigenmodes of the Schrödinger equation~\cite{marra_topological_2024}, we obtain a second eigenmode satisfying the same modified Jackiw-Rebbi equation, but with opposite pseudospin.
These additional modes are not necessarily bounded and do not necessarily satisfy the boundary conditions, even if the original zero mode $\varphi_1(x)$ does.
Finally, a distinct inverse approach is to assume the existence of two distinct zero modes $\varphi_1(x)$ and $\varphi_2(x)$ and derive the spatial dependence of the mass $m(x)$ and of the Dirac velocity $v(x)$, such that the ensuing modified Jackiw-Rebbi equation has these modes as solutions.
These inverse methods generalize straightforwardly to the case of nonhermitian modified Jackiw-Rebbi equations and Schrödinger equations corresponding to nonhermitian Hamiltonians, shedding light on zero energy modes in nonhermitian systems.

Note that, due to the bulk-boundary correspondence~\cite{hatsugai_chern_1993,ryu_topological_2002,teo_topological_2010}, exponentially localized zero modes are always localized at the boundary between topologically inequivalent phases at $x<0$ and $x>0$, i.e., phases with different topological invariants, even if the boundary between the two phases is not sharp (see also Ref.~\cite{marra_topological_2024}).
We recall that for uniform fields $m(x)=m$, $v(x)=v$, one can define a topological invariant $W\in\mathbb{Z}_2$ given by $W=\sgn(v)$ for $m\le0$ and $W=0$ otherwise~\cite{kitaev_unpaired_2001}.
The number of zero modes localized at the boundary is equal to the absolute difference between the topological invariants on the right and on the left of the boundary $\Delta W=|W_L-W_R|$.
In the case of exponentially localized modes, the resulting fields $m(x)$ and $v(x)$ become uniform at large distances $|x|>\infty$ so that the topological invariant is well-defined, at least asymptotically $x\to\pm\infty$.
We anticipate that our results generalize the bulk-boundary correspondence even in cases where the zero modes are not exponentially localized (i.e., faster-than-exponential or slower-than-exponential modes).

Gaussian modes (decaying faster than exponential) of the Schödinger equation correspond to regimes where the mass diverges quadratically for $x\to\pm\infty$.
This scenario describes particles confined in a harmonic trap, which is a typical setup to confine cold atoms in experiments.
Gaussian modes of the modified Jackiw-Rebbi equation correspond to regimes where the mass diverges quadratically for $x\to\pm\infty$, or where the Dirac velocity diverges linearly for $x\to\pm\infty$.
In the case of constant Dirac velocity and quadratically diverging mass, this scenario describes particles in a topological insulator, superconductor, or superfluid in the presence of a harmonic trap, as in the setups considered in Refs.~\cite{zhou_revealing_2013,kheirkhah_majorana_2020}, or nucleons confined by harmonic scalar (or vector) potentials mean-field nuclear shell models~\cite{wen-chao_bound_2002,ti-sheng_pseudospin_2003,alberto_tensor_2005,akcay_dirac_2009,akcay_exact_2009}.
In the case of constant mass and linearly diverging velocity, this scenario may describe particles confined in a Josephson-like junction where the superconducting order parameter is linear in space and changes sign at the junction.
In particular, if the Dirac velocity is uniform, then the mass diverges quadratically, generalizing the harmonic trap scenario, and $\Re(m(x))$ changes sign twice on the real axis.
In this case, for each Gaussian mode with pseudospin $s$ there exists another Gaussian mode with opposite pseudospin $-s$, forming a pair of Gaussian modes with opposite pseudospins.
For real fields, consistently with the bulk-boundary correspondence, there are two zero modes localized at the two points where the mass changes sign, i.e., at the two boundaries between different topological phases with topological invariants $W=0$ and $W=1$, giving $|\Delta W|=1$ modes at each boundary.
Arguably, the faster-than-exponential decay is a consequence of the fact that the mass diverges for $x\to\pm\infty$, such that the topological gap diverges and the localization length $\xi$ goes to zero.
Conversely, if the mass is uniform, the Dirac velocity diverges linearly, and $\Re(v(x))$ changes sign once on the real axis.
In this case, for each Gaussian mode with pseudospin $s$ there exists another mode with the same pseudospin $s$, which however decays polynomially (slower than exponentially), forming a pair of modes with the same pseudospin.
Consistently with the bulk-boundary correspondence, there are two zero modes localized at the points where the Dirac velocity changes sign, i.e., at the boundary between different topological phases with topological invariants $W=-1$ and $W=1$, giving $|\Delta W|=2$ modes at the boundary.

Exponentially localized modes of the Schödinger equation correspond to regimes where the potential exhibits a localized dip and converges to a nonzero constant value for $x\to\pm\infty$.
Exponentially localized modes of the modified Jackiw-Rebbi equation correspond to regimes where the mass and the Dirac velocity converge to a finite and nonzero constant value for $x\to\pm\infty$.
This scenario describes the usual case of a sharp boundary between two distinct topological phases, or the case of a smooth boundary where the two phases are homogeneous (i.e., fields are uniform) at large distances from the boundary (see also Ref.~\cite{marra_topological_2024}).
If the Dirac velocity is uniform, for each mode with pseudospin $s$ there exists another exponentially localized mode with opposite pseudospin $-s$, which is also exponentially localized, under certain conditions.
For real fields, this condition is consistent with the bulk-boundary correspondence, giving only one zero mode if the mass has opposite signs at $x\to\pm\infty$.
Conversely, if the mass is uniform, the Dirac velocity assumes values with opposite signs at $x\to\pm\infty$. 
In this case, for each mode with pseudospin $s$ there exists another mode with the same pseudospin $s$, which is also exponentially localized.
Consistently with the bulk-boundary correspondence, there are two zero modes in this case, since the topological invariants at $x\to\pm\infty$ have opposite signs, giving $|\Delta W|=2$ modes at the boundary.

Polynomially decaying modes (decaying slower than exponential) of the Schödinger equation correspond to regimes where the potential exhibits a localized dip and converges to zero for $x\to\pm\infty$.
Polynomially decaying modes of the modified Jackiw-Rebbi equation correspond to regimes where the mass exhibits a localized dip and converges to zero for $x\to\pm\infty$ or where the Dirac velocity diverges asymptotically (assuming continuous mass and velocity).
These modes can also be referred to as quasi-localized modes since the localization is not exponential but only polynomial.
In particular, if the Dirac velocity is uniform, then the mass converges to zero, and $\Re(m(x))$ changes signs once or three times on the real axis.
In this case, there exists only a single zero mode.
The slower-than-exponential decay is a consequence of the fact that the topological gap goes to zero for $x\to\pm\infty$, such that the localization length $\xi$ diverges.
Hence, one can surmise that zero-energy boundary modes always decay exponentially or faster than exponentially as long as the topological gap does not vanish. 
Conversely, if the mass is uniform, the Dirac velocity diverges linearly, and $\Re(v(x))$ changes sign once on the real axis.
In this case, there exists another mode with the same pseudospin $s$, which, however, decays faster than exponentially.

Hence, in general, the presence and number of zero modes are always determined by the bulk-boundary correspondence.
However, the localization properties of the modes (faster-than, slower-than, or exponential) depend on the detail of the fields at large distances.
For uniform Dirac velocity, faster-than-exponential localization corresponds to divergent masses, exponential localization corresponds to finite masses, and slower-than-exponential localization corresponds to vanishing masses at $x\to\infty$.

Moreover, we found a general condition on the reality of the zero modes.
It is well known that the reality of the modified Jackiw-Rebbi equation and of the Schödinger equation mandates the reality of their eigenmode.
However, the converse is not true:
We find that a real zero mode of the modified Jackiw-Rebbi can correspond to a regime where both $m(x)$ and $v(x)$ are complex.
Perhaps less surprisingly, a real eigenmode of the Schödinger equation can correspond to a regime where both the potential $V(x)$ and the energy $E$ are complex, provided that $V(x)-E$ is real. 

Furthermore, we found examples where two zero modes are localized at a finite distance.
The scenario where two boundary modes appear at a finite distance is typically found in topological insulators and superconductors, where the topological phase necessarily has a finite size (e.g., confined by the physical boundaries of the experimental device).
Finally, we found examples where zero modes are periodic in space, being localized on an equally spaced array of points. 
This regime is realized, e.g., in topological insulators or superconductors in the presence of spatially periodic electric or magnetic fields, where the spatial variations of the fields induce the alternation of topologically trivial and nontrivial segments with zero modes localized at their boundaries, such as the setups investigated in Refs.~\cite{neupert_chain_2010,rex_majorana_2020,marra_1d-majorana_2022,marra_dispersive_2022,marra_majorana_2024}.
The presence of boundary modes at exactly zero energy in these cases is in contrast to the general knowledge that the spatial overlap of multiple localized modes at finite distances lifts the energy away from zero.

\section{The modified Jackiw-Rebbi and associated Schrödinger equation}

As in our previous work~\cite{marra_topological_2024}, we consider the Jackiw-Rebbi equation~\cite{jackiw_solitons_1976} with an additional quadratic term in the momentum given by
\begin{equation}
	\left(\left(
\eta
	p^2+m(x)\right)\tau_z
	+
 {2v}(x)
 p\,\tau_y
 \right)\Psi(x)=E \Psi(x).
\label{eq:H-pwaveC}
\end{equation}
where $\Psi(x)^\dag=[\psi(x)^\dag,\psi(x)]$ is a spinor with $\Psi(x)$ a fermion field, $\tau_{xyz}$ the Pauli matrices with $\tau_\pm=(\tau_x\pm\ii\tau_y)/2$, $p=-\ii\partial$ the momentum operator, and $m(x),v(x)\in\mathbb{C}$.
In the hermitian case and with constant $v(x)=v$, this equation describes: 
a Dirac fermion of mass $m(x)$ if $\eta=0$,
a topological insulator with effective mass $1/2\eta$, chemical potential $\mu(x)=-m(x)$ and spin-orbit coupling $2v(x)$; 
a spinless topological superconductor with effective mass $1/2\eta$, chemical potential $\mu(x)=-m(x)$, and $p$-wave superconducting pairing $\Delta_p(x)=2v(x)$.
The cases where $\Im(m(x))\neq0$ or $\Im(v(x))\neq0$ generalize these equations to the nonhermitian case: 
in particular, $\Im(m(x))\neq0$ describes a topological insulator or superconductor in the presence of a nonhermitian potential,
while $\Im(v(x))\neq0$ the presence of a nonhermitian order parameter in a topological superconductor.
Zero modes $E=0$ are necessarily eigenstates of $\tau_x$ in the form
\begin{equation}\label{eq:spinor}
\Psi(x)\propto
\begin{pmatrix} 
1 \\ 
s
\end{pmatrix}
\varphi(x),
\,\,
\tau_x\Psi(x)=s\Psi(x)
,
\end{equation}
with pseudospin $s=\pm1$ and with $\varphi(x)$ satisfying 
\begin{equation}\label{eq:diffeq}
\varphi''(x)
+{2sv}(x) \varphi'(x)
-
m(x)
\varphi(x)
=0,
\end{equation}
where we set $\eta=1$.
In this case, the zero modes are Majorana modes analogous to the Majorana solutions of the Dirac equations.

We will also consider the Schrödinger equation of a particle with energy $E$ scattering off a potential $V(x)$, given by
\begin{equation}\label{eq:SchrodingerGeneral}
\varphi''(x)
-
\left(V(x)-E\right)
\varphi(x)
=0.
\end{equation}
The solutions of the modified Jackiw-Rebbi equation and of the Schrödinger equation are related by the duality transformation~\cite{marra_topological_2024} $\varphi_\text{SC}(x)=\ee^{s {\int v(x)\dd x}}\varphi_\text{JR}(x)$, assuming the mass $m(x)$, the Dirac velocity $v(x)$, and the potential $V(x)$ are related by the relation $V(x)-E=v(x)^2+sv'(x)+m(x)$.

\section{Inverse method}

The general solutions of the modified Jackiw-Rebbi equation in \cref{eq:diffeq} for constant fields $m(x)=m=\const$ and $v(x)=v=\const$ are well known and are given by the linear combinations $\varphi(x)=A \ee^{(-{sv}-\sqrt{m+{sv}^2})x }+B \ee^{(-{sv} + \sqrt{m+{sv}^2})x }$ for $m\neq0$ and by $\varphi(x)=A+B\ee^{-2 {sv} x}$ for $m=0$.
However, it is not always possible to calculate the explicit analytical form of the solutions of the modified Jackiw-Rebbi equation in \cref{eq:diffeq} for arbitrary choices of the fields $m(x)$ and $v(x)$.
However, the inverse problem is rather straightforward: 
For a given wavefunction $\varphi(x)$, we can reinterpret \cref{eq:diffeq} as an equation for the unknown fields $m(x)$ and $v(x)$.
Hence, by knowing the wavefunction $\varphi(x)$ and the field $v(x)$ explicitly in a given interval $x_\text{min}\le x\le x_\text{max}$, we can calculate 
\begin{equation}\label{eq:inversemu}
m(x)
=
\frac{
\varphi''(x)
+
{2sv}(x) \varphi'(x)
}{
\varphi(x)}
,
\end{equation}
where we take the limit of the right-hand side of the equation above at the nodes of the wavefunction $\varphi(x)=0$.
If the wavefunction has nodes $\varphi(x_i)=0$ for $x_\text{min}\le x_i\le x_\text{max}$, then the field may diverge $|m(x\to x_i)|\to\infty$.
Requiring the convergence of the field $|m(x)|<\infty$ imposes the condition $\varphi''(x)+{2sv}(x) \varphi'(x)\propto\varphi(x)$ in the limit $x\to x_i$ on the wavefunction and on the field $v(x)$.

Conversely, by knowing $\varphi(x)$ and $m(x)$ explicitly,  we can calculate
\begin{equation}\label{eq:inversenu}
s{v}(x) 
=
-\frac{\varphi''(x)-
m(x)
\varphi(x)}{2\varphi'(x)}
,
\end{equation}
where we take the limit of the right-hand side of the equation above at the extrema of the wavefunction $\varphi'(x)=0$.
A renormalizable wavefunction on the real line  $-\infty\le x_i\le \infty$  has necessarily at least one maximum, excluding the trivial case $\varphi(x)=0$.
In general, if the wavefunction has extrema $\varphi'(x_i)=0$ for $x_\text{min}\le x_i\le x_\text{max}$, then the field may diverge $|v(x\to x_i)|\to\infty$.
Requiring the convergence of the field $|v(x)|<\infty$ imposes the condition $\varphi''(x)-m(x)\varphi(x)\propto\varphi'(x)$ in the limit $x\to x_i$ on the wavefunction and on the field $m(x)$.

Analogously, it is not always possible to calculate explicitly the solutions of the Schrödinger equation in \cref{eq:SchrodingerGeneral} for arbitrary choices of the potential $V(x)$.
However, by knowing the wavefunction $\varphi(x)$ explicitly in a given interval $x_\text{min}\le x\le x_\text{max}$, we can calculate 
\begin{equation}\label{eq:inverseV}
V(x)-E
=
\frac{
\varphi''(x)
}{
\varphi(x)}
,
\end{equation}
where we take the limit of the right-hand side of the equation above at the nodes of the wavefunction $\varphi(x)=0$.
If the wavefunction has nodes $\varphi(x_i)=0$ for $x_\text{min}\le x_i\le x_\text{max}$, then the field may diverge $|V(x\to x_i)|\to\infty$.
Requiring the convergence of the field $|V(x)|<\infty$ imposes the condition 
$\varphi''(x)\propto\varphi(x)$ 
in the limit $x\to x_i$ on the wavefunction $\varphi(x)$.
If the wavefunction is real, then the field $V(x)$ is hermitian.

The equations above allow for some simple statements about the reality of the wavefunction and of the fields (hermitianicity).
If the wavefunction of the Jackiw-Rebbi equation is real and one of the fields is hermitian, e.g., $m(x)$ ($v(x)$), this mandates that the other field, e.g., $v(x)$ ($m(x)$) is hermitian as well.
If the wavefunction of the Schrödinger equation is real then the field $V(x)$ is real with real energy $E$ or, alternatively, $V(x)$ is complex with complex energy $E$, but with the difference $V(x)-E$ being real, i.e., with $\Im(V(x))=\Im(E)$.

We also notice that, if we consider a wavefunction $\varphi(x)$ which at the same time is a zero eigenmode of the modified Jackiw-Rebbi equation in \cref{eq:diffeq} with $m(x)$ and $v(x)$ and also an energy eigenmode of the Schrödinger equation in \cref{eq:SchrodingerGeneral} with $V(x)$, then combining \cref{eq:inversemu,eq:inverseV}, one obtains
\begin{align}\label{eq:inverseVmu}
m(x)
=&
V(x) - E
+
2sv(x) 
\frac{\varphi'(x)}{\varphi(x)}
,
\\
s v(x) 
=&
\frac12\left(m(x) - V(x) + E \right)\frac{\varphi(x)}{\varphi'(x)}
,
\end{align}
where $V(x) - E$ is as in \cref{eq:inverseV} and $m(x),v(x)$ as in \cref{eq:inversemu}.

\section{Linearly independent modes with the same pseudospin}

If $\varphi(x)=\varphi_1^s(x)$ is a solution of the (2nd order) differential equation in \cref{eq:diffeq} with pseudospin $s=\pm1$, then by reduction of order, one can immediately get a second linearly independent mode with the same pseudospin $\varphi_2^s(x)$.
Indeed, by writing $\varphi_2^{s}(x)=\theta(x)\varphi_1^s(x)$ one obtains a differential equation   $\varphi_1\theta''+(2\varphi_1'+2s v \varphi_1)\theta'=0$ which is at the first order in $\theta(x)$. 
By integrating one thus obtains
\begin{equation}\label{eq:2ndmode}
\varphi_2^s(x)=\varphi_1^s(x)
\int_0^x
\frac{\ee^{-2s\int_0^y v(z) \dd z}}{(\varphi_1^s(y))^2}\dd y 
,
\end{equation}
up to integration constants.
The two wavefunctions $\varphi_1^s(x)$ and $\varphi_2^s(x)$ are orthogonal if their wronksian  
$W=\varphi_1^s(x) (\varphi_2^s(x))'- (\varphi_1^s(x))' \varphi_2^s(x)=
\ee^{-2s\int_0^x v(y) \dd y}$
is nonzero $W\neq0$.
It is not always possible to find an explicit analytical expression for the mode $\varphi_2^s(x)$ in \cref{eq:2ndmode} in terms of elementary functions.

It should be noted that the mode $\varphi_2^{s}(x)$ does not necessarily decay to zero for $|x|\to\infty$ if $\varphi^{s}(x)$ does.
Assuming $|v(x)|$ bounded, we will see that if the mode $\varphi_1^{s}(x)$ decays to zero for $|x|\to\infty$ faster than exponentially, then $\varphi_2^{s}(x)$ diverges.
Conversely, if the mode $\varphi_1^{s}(x)$ decays slower than exponentially, then $\varphi_2^{s}(x)$ decays to zero on both sides if $s\Re(v_L)<0<s\Re(v_R)$.
Finally, if the mode $\varphi_1^{s}(x)$ decays to zero exponentially, then $\varphi_2^{s}(x)$  can decay to zero exponentially or diverge, depending on the decay rates of the mode for $|x|\to\infty$.
As a general statement, we will see that given a mode $\varphi_1^{s}(x)$ normalizable on the whole real line and decaying to zero for $|x|\to\infty$, the linearly independent mode with the same pseudospin $\varphi_2^{s}(x)$ in \cref{eq:2ndmode} is normalizable on the whole real line and decays to zero for $|x|\to\infty$ only if the real part of $v(x)$ has opposite sign at the asymptote, i.e., only if $v(x\to\mp\infty)=v_{L,R}$ with $\Re(v_L)\Re(v_R)<0$, and only if the pseudospin $s$ is such that $s\Re(v_L)<0<s\Re(v_R)$.
This case is realized, e.g., in a topological insulator or superconductor when the winding number remains nonzero but changes sign on the left and on the right of a domain wall, e.g., being $W=1$ and $W=-1$ for $x<-w$ and $x>w$, respectively, where $w$ is the width of the barrier (see also Ref.~\cite{marra_topological_2024}).

\section{Linearly independent modes with opposite pseudospin}

If $v(x)=v=\const$, there is a straightforward way to obtain a second linearly independent mode  $\varphi(x)=\varphi_1^{-s}(x)$ with opposite pseudospin.
In this case, having $v'(x)=0$, one has that the associated Schrodinger equations obtained for both pseudospins $s=\pm1$ in \cref{eq:SchrodingerGeneral} coincide, since $V(x)-E=v(x)^2+m(x)$, which mandates that
$
\ee^{s v x}\varphi^s(x)
=
\ee^{-s v x}\varphi^{-s}(x)
$,
where $\varphi^s(x)$ are the wavefunctions of the modified Jackiw-Rebbi equation with pseudospin $s=\pm1$.
Alternatively, one can show by direct substitution in \cref{eq:diffeq} that if $v(x)=v=\const$, then if $\varphi^s(x)$ is a solution of the Jackiw-Rebbi equation with pseudospin $s=\pm1$, then $\ee^{2s v x}\varphi^s(x)$ is the solution of the same equation, but with opposite pseudospin.
Hence, the two wavefunctions $\varphi^s(x)$ for $s=\pm1$ of the modified Jackiw-Rebbi equation have a simple relation, given by
\begin{equation}\label{eq:twin}
\varphi^{-s}(x)=\ee^{2s v x}\varphi^{s}(x).
\end{equation}
Hence, if $v(x)=v=\const$, the equation above describes the relation between the mode $\varphi^{s}(x)$ with pseudospin $s$ and the mode $\varphi^{-s}(x)$ with opposite pseudospin $-s$ solutions of the modified Jackiw-Rebbi equation with field $m(x)$ obtained from \cref{eq:inversemu}.
Thus, starting from the mode $\varphi_1^s(x)$, one can obtain the second linearly independent mode $\varphi_2^{s}(x)$ via \cref{eq:2ndmode}, and then the modes with opposite pseudospin $\varphi_1^{-s}(x)$ and $\varphi_2^{-s}(x)$.

If $v'(x)\neq0$, there is no straighforward way to obtain the mode $\varphi^{-s}(x)$ by knowing $\varphi^{s}(x)$ alone.
For instance, one could attempt a reduction of order approach, writing $\varphi^{-s}(x)=\theta(x)\varphi^s(x)$ and obtaining a differential equation for $\theta(x)$.
Unfortunately, the equation obtained is of second order with all coefficients being nonzero, and cannot be reduced to a first-order equation. 
Hence, this approach is no better than solving \cref{eq:diffeq} with $s\to-s$ directly.

The mode $\varphi^{-s}(x)$ does not necessarily decay to zero for $|x|\to\infty$ if $\varphi^{s}(x)$ does.
In general, assuming $|v(x)|$ bounded, we will see that if the mode $\varphi^{s}(x)$ decays to zero for $|x|\to\infty$ faster than exponentially, then $\varphi^{-s}(x)$ also decays faster than exponentially.
Conversely, if the mode $\varphi^{s}(x)$ decays slower than exponentially, then $\varphi_2^{s}(x)$ decays to zero only if $s\Re(v_{L})\ge0$ and if $s\Re(v_{R})\le0$.
Finally, if the mode $\varphi^{s}(x)$ decays to zero exponentially, then $\varphi^{-s}(x)$ can decay to zero exponentially or diverge depending on the decay rates of the mode for $|x|\to\infty$.

\section{Inverse method with two modes}

As we have seen, it is not always possible to obtain the mode with opposite pseudospin $\varphi^{-s}(x)$ from the mode $\varphi^s(x)$, except for the simple case where $v(x)=v=\const$, and it is not always possible to obtain an analytical expression of the second linearly independent mode $\varphi_2^{s}(x)$ from a given mode $\varphi_1^{s}(x)$.
However, one can reframe this problem as an inverse problem as well.
In general, given two modes $\varphi_1(x)=\varphi_1^{s_1}(x)$ and $\varphi_2(x)=\varphi_2^{s_2}(x)$ satysfing the equation in \cref{eq:diffeq} with pseudospin $s=s_1$ and $s=s_2$ respectively, one can obtain the fields $m(x)$ and $v(x)$ from \cref{eq:inversemu,eq:inversenu}, which combined give
\begin{align}
m(x)=&
\frac
{\varphi _1'' \varphi _2' -s_1 s_2 \varphi _1' \varphi _2''}
{\varphi _1 \varphi _2'-s_1 s_2 \varphi _1' \varphi _2}
,
\label{eq:inversemu2modes}
\\
v(x)=&
\frac
{s_2 \left(\varphi _1''\varphi _2  - \varphi _1 \varphi _2'' \right)}
{2 \left(\varphi _1 \varphi _2'-s_1 s_2 \varphi _1' \varphi _2 \right)}
,
\label{eq:inversenu2modes}
\end{align}
where we take the limit of the right-hand side of the equations above at the zeros of the denominators.

\section{Analytical solutions}

In the rest of this paper, we will employ this recipe to study several examples of modes with wavefunctions $\varphi_1^s(x)$ defined on the infinite interval $-\infty< x<\infty$ and calculate:
i) the associated field $V(x)$ which satisfies \cref{eq:SchrodingerGeneral} via \cref{eq:inverseV},
ii) the associated field $m(x)$ which satisfies \cref{eq:diffeq} 
with $v(x)=v=\const$ 
via \cref{eq:inversemu} as a function of $v$
iii) the associated field $v(x)$ which satisfies \cref{eq:diffeq} with $m(x)=m=\const$ 
via \cref{eq:inversenu} as a function of $m$.
iv) the associated mode with same pseudospin $\varphi_2^s(x)$ via \cref{eq:2ndmode} and with opposite pseudospin $\varphi_{1}^{-s}(x)$ via \cref{eq:twin} when $v(x)=v=\const$, if bounded.
Moreover, we will consider a pair of modes with wavefunctions $\varphi_1^{s_1}(x+a)$ and $\varphi_2^{s_2}(x-a)$ with $s_1s_2=\pm1$ and calculate
v) the associated fields $m(x)$ and $v(x)$ via \cref{eq:inversemu2modes,eq:inversenu2modes}.
We will consider modes with exponential decay, Gaussian decay (faster-than-exponential), and polynomial decay (slower-than-exponential) with real wavefunctions or complex wavefunctions with nonuniform complex phases.
We will also show examples of multilocational periodic modes, i.e., having wavefunctions with probability density peaking on an equally spaced array of points.

As a simple example to illustrate the procedures, by considering plain waves
$
\varphi(x) = \ee^{\ii \alpha x}
$,
with $\alpha>0$, one obtains that
the associated field $V(x)$ satisfying \cref{eq:SchrodingerGeneral} calculated via \cref{eq:inverseV} is given by
$
V(x) - E = - \alpha^2
$,
the associated field $m(x)$ satisfying \cref{eq:diffeq} calculated via \cref{eq:inversemu} is given by
$
m(x) = V(x) - E + 2s  \ii \alpha v(x) 
$,
and the associated field $v(x)$ satisfying \cref{eq:diffeq} calculated via \cref{eq:inversenu} are given by
$
s v(x) = - \frac{\ii  }{2\alpha} \left(\alpha^2 +  m(x)\right)
$.
This mandates that plane wave solutions of the Jackiw-Rebbi equation with $v(x)\neq0$ can be obtained only in the nonhermitian case.
Moreover, the second linearly independent mode with the same pseudospin obtained from \cref{eq:2ndmode} and with $m(x)=m=\const$ and $v(x)$ as in the previous equation is given by $\varphi(x)=\ee^{\frac{\ii m x}{\alpha }}$.
The corresponding modes with opposite pseudospin obtained when $v(x)=v=\const$ from \cref{eq:twin} are $\ee^{(2sv+\ii \alpha) x}$ and $\ee^{-\ii\alpha x}$ which are again normalizable only if $\Re(v)=0$.

\begin{figure}[tbp]
   \centering\includegraphics{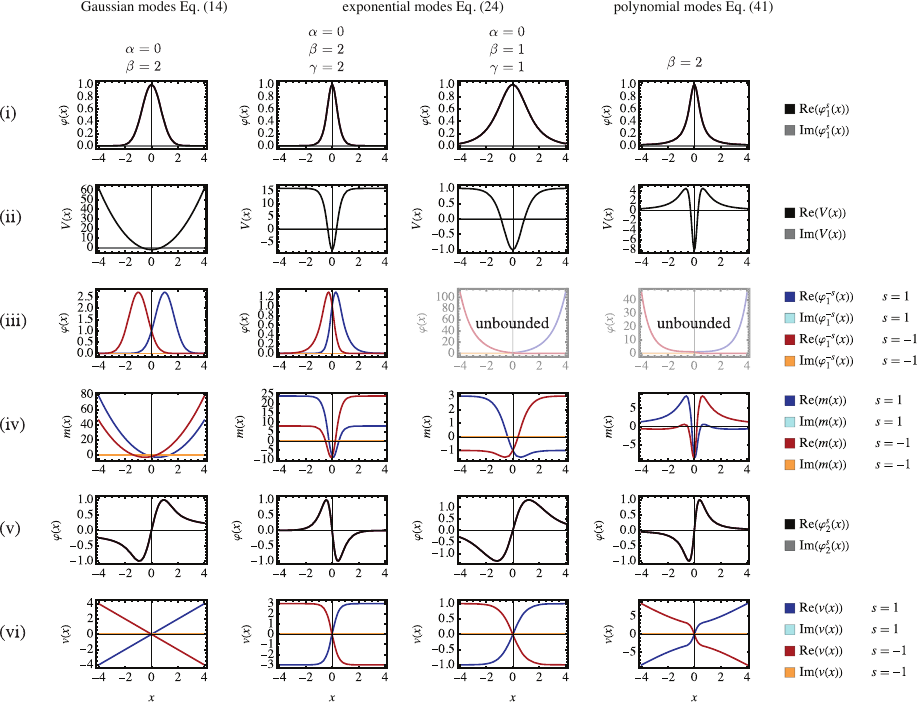}
   \caption{
Modes with real wavefunction and Gaussian, exponential, or polynomial decay on the infinite interval $-\infty< x<\infty$ and associated fields for some choice of the parameters.
(i) wavefunction $\varphi_1^s(x)=\varphi(x)$ as in \cref{eq:gauss,eq:sech,eq:witch},
(ii) associated field $V(x)$ satisfying \cref{eq:SchrodingerGeneral},
(iii) the wavefunction $\varphi_1^{-s}(x)$ of the associated mode in \cref{eq:twin} with opposite pseudospin (if bounded) and 
(iv) the associated field $m(x)$ for $s=\pm1$ satisfying \cref{eq:diffeq} for a constant field $v(x)=v=\const$ (where $v=1$),
(v) the wavefunction $\varphi_2^{s}(x)$ of the associated mode in \cref{eq:2ndmode} with same pseudospin and 
(vi) associated fields $v(x)$ for $s=\pm1$ satisfying \cref{eq:diffeq} 
for a constant field $m(x)=m=\const$ where $m =m^*$, such that the field $v(x)$ is continuous and bounded ($m^*=-\beta$, $(\alpha^2-\beta^2\gamma^2)/\gamma$, and $-2\beta^2$ respectively for Gaussian, exponentially decaying, and polynomially decaying modes).
For Gaussian modes, the fields $V(x)$ and $m(x)$ (for $v(x)=v=\const$) diverge quadratically, with the field $m(x)$ changing sign twice on the real axis, while the field $v(x)$ (for $m(x)=m=\const$) diverges linearly, changing sign only once.
In this case, both the associated mode with opposite pseudospin for $v(x)=v=\const$
and the associated mode with same pseudospin for $m(x)=m=\const$ with $m=m^*$ are bounded and thus normalizable.
For exponentially decaying modes, the fields $V(x)$, $m(x)$, and $v(x)$ converge to finite values for $x\to\pm\infty$.
In this case, the associated mode with the opposite pseudospin is bounded only for 
$\Re(\beta\gamma+\alpha)>2s \Re(v)>-\Re(\beta\gamma-\alpha)$ 
while the associated mode with the same pseudospin is bounded only if
$2s\Re(v_{L})<-\Re(\beta\gamma+\alpha)$ and
$2s\Re(v_{R})>\Re(\beta\gamma-\alpha)$.
For polynomially decaying modes, the fields $V(x)$ and $m(x)$ (for $v(x)=v=\const$) converge to zero, with the field $m(x)$ changing sign twice on the real axis, while the field $v(x)$ (for $m(x)=m=\const$) diverges linearly, changing sign only once.
In this case, the associated mode with opposite pseudospin is bounded only in the limiting case where $\Re(v)=0$,
while the associated mode with the same pseudospin is bounded as long as $s\Re(v_R)>0>s\Re(v_L)$.
}
   \label{fig:real}
\end{figure}

\begin{figure}[tbp]
   \centering\includegraphics{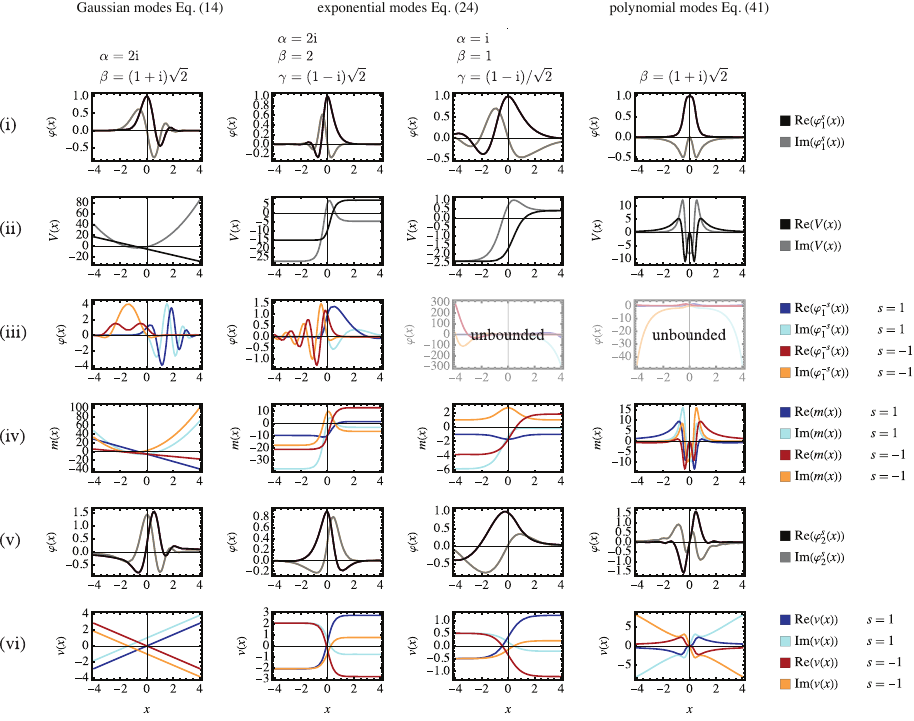}
   \caption{
Same as in \cref{fig:real} but for modes with nonuniform complex phases and nonhermitian fields.
}
   \label{fig:complex}
\end{figure}

\subsection{Gaussian-like modes}

We consider modes with Gaussian decay (faster-than-exponential) given by the wavefunction
\begin{equation}\label{eq:gauss}
\varphi(x) = \ee^{ - \frac{\beta x^2}{2} - \alpha x}
,
\end{equation}
with $\Re(\beta)>0$, having norm 
$
\Vert\varphi(x)\Vert^2 = 
\ee^{\frac{\Re(\alpha )^2}{\Re(\beta )}}
\sqrt{\frac{\pi}{\Re(\beta )}}
$.
The wavefunction has a Gaussian bell-like shape and decays faster than exponentially for $|x|\to\infty$.

The associated field $V(x)$ satisfying \cref{eq:SchrodingerGeneral} calculated via \cref{eq:inverseV} is given by
\begin{equation}
V(x) - E = 
(\beta x + \alpha)^2 - \beta
,
\end{equation}
which is a polynomial of the second order in $x$.
The associated field $m(x)$ satisfying \cref{eq:diffeq} calculated via \cref{eq:inversemu} is given by
\begin{align}
m(x) &= V(x) - E - 2s \left(\beta x + \alpha \right) v(x)
\nonumber\\&=
(\beta x + \alpha)^2 - \beta
- 2s \left(\beta x + \alpha \right) v(x)
,
\label{eq:mgauss}
\end{align}
which is a polynomial of the second order in $x$ if $v(x)=v=\const$ (or if $v(x)=a+b x$).
The fields $V(x)$ and $m(x)$ are harmonically shaped potentials that diverge for $|x|\to\infty$.
Note that in the real case $\alpha,\beta,v\in\mathbb{R}$, the field $m(x)$ crosses zero two times since the discriminant of the quadratic equation $m(x)=0$ is always positive.\footnote{Quadratic equations with real coefficients have two distinct roots when the discriminant is positive.
The discriminant of the quadratic equation $m(x)=0$ is $v^2+\beta>0$.
}
The associated field $v(x)$ satisfying \cref{eq:diffeq} calculated via \cref{eq:inversenu} is given by
\begin{equation}\label{eq:vgauss}
s v(x)= \frac{\beta x+\alpha}{2} - \frac{\beta + m(x)}{2 (\beta x + \alpha) }
,
\end{equation}
which is a rational function in $x$ if $m(x)=m=\const$ (or $m(x)$ is a polynomial in $x$).
This field diverges for $|x|\to\infty$ 
and diverge for $x\to x^*$ where 
$
x^* = - \frac{\alpha }{\beta }
$
in the case that $x^*\in\mathbb{R}$ and 
$
m(x^*) \neq m^*$ with
$
m^* = - \beta
$.
In the case where 
$m(x) = m^*$ 
the last term in \cref{eq:vgauss} cancels out and thus becomes
\begin{equation}\label{eq:vstargauss}
s v(x)=  \frac{\beta x+\alpha}{2}
,
\end{equation}
which is a polynomial of the first order in $x$, and thus, it is continuous but not bounded.
If $x^*\in\mathbb{R}$ and $m(x)$ is not constant but $m(x^*) = m^*$, then the field converges if $(m(x)-m(x^*))/(x-x^*)$ converges for $x\to x^*$.
The fields $V(x)$ and $m(x)$ are polynomials of the second order in $x$.
The field $v(x)$ is a polynomial of the first order or a rational function of $x$.

For $v(x)=v=\const$, the associated mode with opposite pseudospin in \cref{eq:twin} given by
\begin{equation}
\varphi_1^{-s}(x)=\ee^{ - \frac{\beta x^2}{2} + (2s v- \alpha) x},
\end{equation}
decays to zero for $|x|\to\infty$ faster than exponentially, and it is thus normalizable.
Moreover, the associated mode with the same pseudospin in \cref{eq:2ndmode} 
which is $\varphi_2^{s}(x)\propto \ee^{-\frac{\beta  x^2}{2}-\alpha  x} \erfi\left(\frac{\beta  x+\alpha -s v}{\sqrt{\beta }}\right)$ (where $\erfi(x)=\frac{2}{\sqrt{\pi}}\int_0^x\ee^{y^2}\dd y$) diverges in this case.
Since $\varphi_2^{s}(x)$ diverges faster than exponentially, also the associated mode $\varphi_2^{-s}(x)$ diverges.

Conversely, assuming $v(x)\approx v_{L,R}$ respectively for  $x\to\mp\infty$, then the associated mode with same pseudospin in \cref{eq:2ndmode} gives 
$\varphi_2^{s}(x)\propto 
\erfi\left(\frac{\beta x+\alpha-s v_{L,R}}{\sqrt{\beta}}\right)$ 
asymptotically, and therefore diverges.
However, for $m(x)=m=\const$ with $m=m^*$ and $v(x)$ as in \cref{eq:vstargauss}, one has that $|v(x)|\to\infty$ for  $x\to\mp\infty$.
In this case the associated mode with the same pseudospin in  \cref{eq:2ndmode} given by
\begin{equation}
\varphi_2^{s}(x)=\ee^{-\frac{\beta  x^2}{2}-\alpha  x} \erfi\left(\frac{\beta  x+\alpha }{\sqrt{2 \beta }}\right)
,
\end{equation}
decays to zero for $|x|\to\infty$ 
polynomially as $\sim1/x$, 
and it is thus normalizable.

The Gaussian modes and the associated fields are shown in the first column of \cref{fig:real} and the first column of \cref{fig:complex} for the cases of real and complex wavefunctions, respectively.

Since $m(x)$ and $v(x)$ in \cref{eq:mgauss,eq:vgauss} are polynomials of the second and first order in $x$, respectively, one may attempt to write a general solution in the form of \cref{eq:gauss} corresponding to an arbitrary choice of the fields $m(x)=m_0+m_1 x+m_2 x^2$ and $v(x)=v_0+v_1 x$ by equating the coefficients of the field $m(x)$ in \cref{eq:mgauss}.
By doing this, one obtains a set of three equations
\begin{align}
 m_0 &=\alpha ^2-\beta -2 \alpha  s v_0,\\
 m_1 &=2 \alpha  \beta -2 \beta  s v_0-2 \alpha  s v_1,\\
 m_2 &= \beta ^2-2 \beta  s v_1,
\end{align}
in the two variables $\alpha$ and $\beta$, which cannot be solved for arbitrary choices of the parameters $m_{0,1,2}$ and $v_{1,2}$.
In other words, the wavefunctions in the form \cref{eq:gauss} are only special cases and do not describe the solutions of the \cref{eq:diffeq} for all choices of the fields in the form $m(x)=m_0+m_1 x+m_2 x^2$ and $v(x)=v_0+v_1 x$.
The general analytical solutions for these fields will be reported elsewhere.

Hence, Gaussian modes of the Schödinger equation correspond to cases where the mass diverges quadratically for $x\to\pm\infty$, while
Gaussian modes of the modified Jackiw-Rebbi equation correspond to cases where the mass or the Dirac velocity diverges asymptotically for $x\to\pm\infty$.
If the Dirac velocity is uniform, the mass diverges quadratically, and $\Re(m(x))$ changes sign twice on the real axis.
In this case, for each mode with pseudospin $s$ there exists another mode with opposite pseudospin $-s$.
For real fields, there are two zero modes localized at the two points where the mass changes sign, consistently with the bulk-boundary correspondence.
Conversely, if the mass is uniform, the Dirac velocity diverges linearly, and $\Re(v(x))$ changes sign once on the real axis.
In this case, for each mode with pseudospin $s$ there exists another mode with the same pseudospin $s$, which, however, decays polynomially.
Hence, there are two zero modes localized at the points where the Dirac velocity changes sign, consistently with the bulk-boundary correspondence.

\subsection{Exponentially-decaying modes}

Here we consider normalizable and exponentially decaying modes on the infinite interval $-\infty< x<\infty$ given by the wavefunction
\begin{equation}\label{eq:sech}
\varphi(x) = \frac{\ee^{-\alpha x}}{\cosh^{\gamma}(\beta x)}
,
\end{equation}
with $\Re(\beta\gamma\pm\alpha)>0$, having norm 
$
\Vert\varphi(x)\Vert^2 <\infty
$.
In the limiting case $\Re(\alpha)=\Re(\beta\gamma)=0$ instead, this wavefunction describes a mode with constant probability density $|\varphi(x)|^2=1$ on the whole real line.
Otherwise, the wavefunction decays exponentially as $\ee^{\pm\lambda_{L,R} x}$ for $x\to\mp\infty$, where $\lambda_{L,R}=\Re(\beta\gamma\mp\alpha)>0$.
Note that the mode defined above in \cref{eq:sech} is a special case of the mode in Eq. (5) and (7) of Ref.~\cite{marra_topological_2024} when $a_{1,2}=0$ or $b_{1,2}=0$.

The associated field $V(x)$ satisfying \cref{eq:SchrodingerGeneral} calculated via \cref{eq:inverseV} is given by
\begin{equation}
V(x) - E = \beta^2\gamma\left( \gamma +  1 \right) y^2  + 2 \alpha \beta \gamma y + \alpha^2 - \beta^2 \gamma
,
\end{equation}
which is a polynomial of the second order in $y=\tanh(\beta x)$.
The associated field $m(x)$ satisfying \cref{eq:diffeq} calculated via \cref{eq:inversemu} is given by
\begin{align}\label{eq:msech}
m(x) &= V(x) - E - 2s \left(\beta \gamma y + \alpha \right) v(x)
\nonumber\\&=
\beta^2\gamma\left( \gamma +  1 \right) y^2  + 2 \alpha \beta \gamma y + \alpha^2 - \beta^2 \gamma
- 2s \left(\beta \gamma y + \alpha \right) v(x)
,
\end{align}
which is a polynomial of the second order in $y$ if $v(x)=v=\const$ (or if $v(x)=a+b y$).
These two fields are continuous and bounded $|V(x)|\le\text{max}|V(x)|<\infty$, $|m(x)|\le\text{max}|m(x)|<\infty$ and converge to finite values for $|x|\to\infty$ (assuming $v(x)$ bounded).
For real fields, $m(x)$ has opposite signs at $x\to\pm\infty$ if $(2 s v-\alpha-\beta\gamma)(2 s v-\alpha+\beta\gamma)>0$.
The associated field $v(x)$ satisfying \cref{eq:diffeq} calculated via \cref{eq:inversenu} is given by
\begin{equation}\label{eq:vsech}
s v(x)= 
\left(\frac{\beta (\gamma +1) }{2} \right) y 
+ \frac{\alpha (\gamma-1)}{2\gamma}
+ \frac{\alpha^2 - \beta^2 \gamma^2 - \gamma m(x)}{2 \gamma(\beta \gamma y + \alpha ) }
,
\end{equation}
which is a rational function in $y$ if $m(x)=m=\const$ (or $m(x)$ is a polynomial in $y$).
This field converges to zero for $|x|\to\infty$ 
but diverges for $x\to x^*$ with $x^*$ given by $\tanh(\beta x^*)=y^*$ where 
$
y^* = - {\alpha }/{\beta \gamma }
$
in the case that $y^*\in\mathbb{R}$ and 
$
m(x^*) \neq m^*$ with
$
m^* = {(\alpha^2- \beta^2 \gamma^2)}/{\gamma } 
$.
In the case where 
$m(x) = m^*$ 
the last term in \cref{eq:vsech} cancels out and thus becomes
\begin{equation}\label{eq:vstarsech}
s v(x)= 
\left(\frac{\beta (\gamma +1) }{2} \right) y 
+ \frac{\alpha (\gamma-1)}{2\gamma}
,
\end{equation}
which is a polynomial in the first order in $y$, continuous and bounded.
If $y^*\in\mathbb{R}$ and $m(x)$ is not constant but $m(x^*) = m^*$, then the field converges if $(m(x)-m(x^*))/(y-y^*)$ converges for $x\to x^*$.

Excluding the case where $v(x)$ is a rational function, the modes considered here are special cases of the localized modes described in Ref.~\onlinecite{marra_topological_2024}, which describe the general solutions corresponding to the fields $m(x)$ and $v(x)$ expanded up to the second order and $v(x)$ up to the first order in $y$, respectively.

For $v(x)=v=\const$, the associated mode with opposite pseudospin in \cref{eq:twin} given by 
\begin{equation}
\varphi_1^{-s}(x)=\frac{\ee^{(2s v- \alpha) x}}{\cosh^{\gamma}(\beta x)},
\end{equation}
decays to zero if $\lambda_{L,R}\pm2s \Re(v)>0$, i.e., 
if the condition $\Re(\beta\gamma+\alpha)>2s \Re(v)>-\Re(\beta\gamma-\alpha)$ is satisfied.
For real fields, this mandates that $m(x)$ has the same sign at $x\to\pm\infty$.
Moreover, the associated mode with the same pseudospin $\varphi_2^{s}(x)$ in \cref{eq:2ndmode} which gives asymptotically $\varphi_2^{s}(x)\propto \ee^{(\mp\lambda_{L,R}-2s \Re(v))x}$ diverges, as well as the associated mode $\varphi_2^{-s}(x)\propto\ee^{\mp\lambda_{L,R}x}$.

Conversely, assuming $v(x)\approx v_{L,R}$ respectively for $x\to\mp\infty$, 
e.g., with $s v_{L,R}=\mp{\beta (\gamma +1) }/{2}   {\alpha (\gamma-1)}/{2\gamma} $ as in \cref{eq:vstarsech},
then the associated mode with the same pseudospin \cref{eq:2ndmode} gives 
asymptotically $\varphi_2^{s}(x)\propto \ee^{(\mp\lambda_{L,R}-2s \Re(v_{L,R}))x}$ respectively for $x\to\mp\infty$,
which decays to zero if $-\lambda_{L,R}\mp2s \Re(v_{L,R})>0$, i.e., if
$2s\Re(v_{L})<-\Re(\beta\gamma+\alpha)$ and
$2s\Re(v_{R})>\Re(\beta\gamma-\alpha)$, 
which also mandates $s\Re(v_R)>0>s\Re(v_L)$.
Moreover, the other associated modes $\varphi_1^{-s}(x)\propto \ee^{(\pm\lambda_{L,R}+2s \Re(v_{L,R}))x}$
and $\varphi_2^{-s}(x)\propto \ee^{\mp\lambda_{L,R}x}$ diverge and are not normalizable.
Specifically, 
for $m(x)=m=\const$ with $m=m^*$ and $v(x)$ as in \cref{eq:vstarsech} with $\gamma\neq0$, then \cref{eq:2ndmode} gives 
\begin{equation} 
\varphi_2^s(x)=\frac{\ee^{-\alpha x}}{\cosh^{\gamma}(\beta x)}
\int_0^x \,
{\cosh^{\gamma-1}(\beta y)}
{\ee^{
\frac{\alpha (\gamma+1)}{\gamma} y
}
}
\dd y
,
\end{equation}
which yields
\begin{equation} 
\varphi_2^s(x)=\frac{x}{\cosh(\beta x)}
,
\end{equation}
for $\gamma=1$ and $\alpha=0$ and 
\begin{equation} 
\varphi_2^s(x)=
{{\ee^{\frac{\alpha +\beta\gamma}{\gamma } x}} }
\, _2F_1\left(1,\frac{ (\alpha +\beta  \gamma )(\gamma +1)}{2 \beta  \gamma },
\frac{\alpha  (\gamma +1)-\beta  \gamma  (\gamma -1)}{2 \beta  \gamma }+1
,-\ee^{2 x \beta }\right)
,
\end{equation}
otherwise.

The exponentially decaying modes and the associated fields are shown in the second column of \cref{fig:real} and the second column of \cref{fig:complex} for the cases of real and complex wavefunctions, respectively.

Since $m(x)$ and $v(x)$ in \cref{eq:msech,eq:vsech} are polynomials of the second and first order, respectively, in $y=\tanh(\beta x)$ for a given $\beta$, one may again attempt to write a general solution in the form of \cref{eq:sech} corresponding to an arbitrary choice of the fields $m(x)=m_0+m_1 y+m_2 y^2$ and 
$v(x)=v_0+v_1 y$ 
by equating the coefficients of the field $m(x)$ in \cref{eq:msech}.
By doing this, one obtains again a set of three equations
\begin{align}
 m_0 &=\alpha ^2-\beta ^2 \gamma -2 \alpha  s v_0,\\
 m_1 &=2 \alpha  \beta  \gamma -2 s (\beta  \gamma  v_0+2 \alpha  v_1),\\
 m_2 &=\beta ^2 \gamma  (\gamma +1)-2 \beta  \gamma  s v_1,
\end{align}
in the two variables $\alpha$ and $\gamma$ (since $\beta$ is determined by the relation $y=\tanh(\beta x)$), which cannot be solved for arbitrary choices of the parameters $m_{0,1,2}$ and $v_{1,2}$.
Again one can conclude that the wavefunctions in the form \cref{eq:sech} are only special cases and do not describe the solutions of the \cref{eq:diffeq} for all choices of the fields in the form $m(x)=m_0+m_1 y+m_2 y^2$ and $v(x)=v_0+v_1 y$.

Hence, exponentially localized modes of the Schödinger equation correspond to cases where the potential exhibits a localized dip and converges to a nonzero constant value for $x\to\pm\infty$, while
exponentially localized modes of the modified Jackiw-Rebbi equation correspond to cases where the mass and the Dirac velocity converge to a finite and nonzero constant value for $x\to\pm\infty$.
If the Dirac velocity is uniform, for each mode with pseudospin $s$ there exists another mode with opposite pseudospin $-s$.
For real fields, this condition is consistent with the bulk-boundary correspondence, giving only one zero mode if the mass has opposite signs at $x\to\pm\infty$.
Conversely, if the mass is uniform, the Dirac velocity assumes opposite signs at $x\to\pm\infty$. 
In this case, for each mode with pseudospin $s$ there exists another mode with same pseudospin $s$.
Hence, there are two zero modes, consistently with the bulk-boundary correspondence.

\subsection{Exponentially-decaying modes and Pöschl–Teller potential}

The exponentially-decaying modes considered in \cref{eq:sech} with $\alpha=0$ are related to the modes of the Schrôdinger equation with Pöschl–Teller potentials.
In this case one has 
\begin{equation}\label{eq:sech-PT}
\varphi(x) = {\cosh^{-\gamma}(\beta x)}
\end{equation}
and
the associated field $V(x)$ satisfying \cref{eq:SchrodingerGeneral} calculated via \cref{eq:inverseV} is given by
\begin{equation}
V(x) - E = 
\beta^2\left(
\gamma(\gamma + 1) y^2  - \gamma
\right)
,
\end{equation}
which coincides with the case of a Pöschl–Teller potential 
$V(x)=
\beta^2\left(
\gamma(\gamma + 1) (\tanh^2(\beta x)  - 1)
\right)
$
and energy of the groundstate $E=-\beta^2\gamma^2$.
The wavefunction of the Schrôdinger equation with Pöschl–Teller potential is given in terms of the Legendre functions $P_a^b(y)$, which in this case give
\begin{equation}
\varphi_\text{PT}(x)=P_\gamma^\gamma(y)\propto
\left(\frac{1-y}{1+y}\right)^{\gamma/2}
{{\,}_2F_1}\left(-\gamma,\gamma+1,\gamma+1,\frac{1-y}2\right)
\propto
\left(1-y^2\right)^{\gamma/2}
,
\end{equation}
up to constant terms depending only on $\gamma$,
where ${{\,}_2F_1}$ is the hypergeometric function, and  $y=\tanh(\beta x)$ as usual.
The fact that this solution coincides with \cref{eq:sech-PT} follows from the identity
$\cosh^2(x)=1-\tanh ^2(x)$.
The associated field $m(x)$ satisfying \cref{eq:diffeq} calculated via \cref{eq:inversemu} is given by
\begin{align}
m(x) &= V(x) - E - 2s  \beta \gamma y v(x)
\nonumber\\&=
\beta^2\left(
\gamma(\gamma + 1) (\tanh^2(\beta x)  - 1)
\right)
- 2s  \beta \gamma y v(x),
\end{align}
while 
the associated field $v(x)$ satisfying \cref{eq:diffeq} calculated via \cref{eq:inversenu} is given by
\begin{equation}\label{eq:vsech-PT}
s v(x)= \left(\frac{\beta (\gamma + 1)}{2} \right) y - \frac{ \beta^2 \gamma^2 + \gamma m(x)}{2 \beta\gamma^2 y }.
\end{equation}
This field converges to zero for $|x|\to\infty$ 
but diverges for $x\to 0$ 
in the case that 
$
m(x^*) \neq m^*$ with
$
m^* = - \beta^2 \gamma
$.
In the case where 
$m(x) = m^*$ 
the last term in \cref{eq:vsech-PT} cancels out.

\subsection{Polynomially decaying modes}

We will now consider some examples of normalizable but not exponentially decaying modes.
For instance, the wavefunction
\begin{equation}\label{eq:witch}
\varphi(x) = \frac{1}{(\beta^2 x^2+1)^{n/2}}
,
\end{equation}
with $\Re(\beta^2)>0$ and $n>0$, 
which decays polynomially as $\propto1/|x|^n$ and
with finite norm (e.g.,  
$
\Vert\varphi(x)\Vert^2 = {\pi }/{2 \Re(\beta )}
$ for $n=2$).
For the mode in \cref{eq:witch} with $n=2$, the associated field $V(x)$ satisfying \cref{eq:SchrodingerGeneral} calculated via \cref{eq:inverseV} is given by
\begin{equation}
V(x) - E = \frac{2 \beta^2(3 \beta^2 x^2 - 1)}{(\beta^2 x^2+1)^2}
,
\end{equation}
which is a rational function with the numerator being a polynomial of the second order in $x$.
The associated field $m(x)$ satisfying \cref{eq:diffeq} calculated via \cref{eq:inversemu} is given by
\begin{align}
m(x) &= V(x) - E 
- s 
\frac{4 \beta ^2 x  v(x)}{\beta ^2 x^2+1}
\nonumber\\&=
\frac{2 \beta^2(3 \beta^2 x^2 - 1)}{(\beta^2 x^2+1)^2}
- s 
\frac{4 \beta ^2 x  v(x)}{\beta ^2 x^2+1}
,
\end{align}
which is a rational function with the numerator being a polynomial of the third order in $x$ if $v(x)=v=\const$.
These two fields are continuous and bounded and decay to zero for $|x|\to\infty$.
Note that in the real case $\alpha,\beta,v\in\mathbb{R}$, the field $m(x)$ crosses zero once or three times, depending on the sign of the discriminant of the cubic equation describing $m(x)$.\footnote{Cubic equations with real coefficients have three distinct real roots or only one real root depending on the sign of the discriminant.}
The associated field $v(x)$ satisfying \cref{eq:diffeq} calculated via \cref{eq:inversenu} is given by
\begin{equation}\label{eq:vwitch}
s v(x) = 
\frac{6 \beta ^4 x^2-2 \beta ^2
-\left(\beta ^2 x^2+1\right)^2m(x) 
}{4 \beta ^2 x \left(\beta ^2 x^2+1\right)
}
,
\end{equation}
which is a rational function if $m(x)$ is a polynomial in $x$.
This field diverges for $|x|\to\infty$ if $m(x)\neq0$
and diverges for $x\to 0$ 
in the case that 
$m(0) \neq m^*$ with
$
m^* = - 2 \beta^2
$.
In the case where 
$m(x) = m^*$,
\cref{eq:vwitch} becomes
\begin{equation}\label{eq:vstarwitch}
s v(x)= \frac{\beta^2({\beta^2 x^2} + {5 }) x}{2(\beta^2 x^2+1 )} 
,
\end{equation}
which is continuous but not bounded in this case and becomes linear $v(x)\propto |x|$ asymptotically for $|x|\to\infty$.
If $m(x)$ is not constant but $m(0) = m^*$, then the field converges if $(m(x)-m(0))/x$ converges for $x\to 0$.

For $v(x)=v=\const$, the associated mode with opposite pseudospin in \cref{eq:twin} given by
\begin{equation}
\varphi_1^{-s}(x)=\frac{\ee^{2s v x}}{(\beta^2 x^2+1)^{n/2}},
\end{equation}
decays to zero for $|x|\to\infty$ and it is thus normalizable only in the case $\Re(v)=0$.
Moreover, the associated mode with the same pseudospin in \cref{eq:2ndmode} 
which is $\varphi_2^{s}(x)\propto x^n\ee^{-s v x}$ at the leading order, diverges for $|x|\to\infty$, as well as the associated mode $\varphi_2^{-s}(x)\propto x^n\ee^{s v x}$.

Conversely, the associated mode with the same pseudospin $\varphi_2^{s}(x)$ in \cref{eq:2ndmode} decays to zero if $s\Re(v_L)<0<s\Re(v_R)$, i.e., if $\Re(v_L)\Re(v_R)<0$ and if the pseudospin is such that $s=\sgn(\Re(v_R))=-\sgn(\Re(v_L))$.
Indeed, assuming $v(x)\approx v_{L,R}$ respectively for $x\to\mp\infty$, then \cref{eq:2ndmode} gives $\varphi_2^{s}(x)\propto x^n\ee^{-s v_{L,R} x}$ at the leading order, which decays to zero for $|x|\to\infty$ if $s\Re(v_R)>0>s\Re(v_L)$.
However, for polynomial modes with $n=2$ and for $m(x)=m=\const$ with $m=m^*$ and $v(x)$ as in \cref{eq:vstarwitch}, 
one has that $|v(x)|\to\infty$ for $|x|\to\infty$.
In this case \cref{eq:2ndmode} gives
\begin{equation}
\varphi_2^{s}(x)=\frac{\erf\left(\frac{\beta  x}{\sqrt{2}}\right)}{\beta ^2 x^2+1},
\end{equation}
which decays faster than exponential, where $\erf(x)=\frac{2}{\sqrt{\pi }}\int _0^x e^{-y^2}\dd y $.

The polynomially decaying modes and the associated fields are shown in the third and fourth columns of \cref{fig:real} and the third and fourth columns of \cref{fig:complex} for the cases of real and complex wavefunctions, respectively.

\subsection{Two modes}

\begin{figure}[tbp]
   \centering\includegraphics{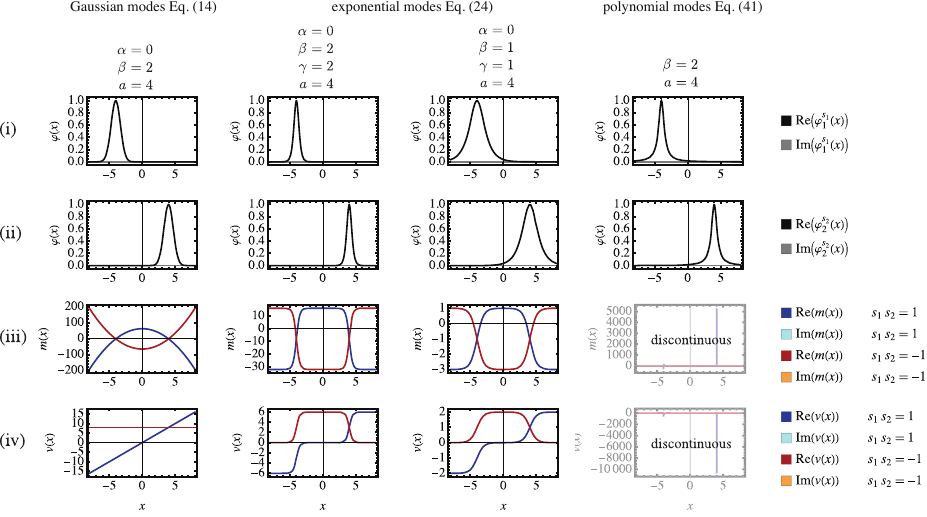}
   \caption{
Couples of modes with real wavefunction and Gaussian, exponential, or polynomial decay and associated fields for some choice of the parameters.
(i) wavefunction $\varphi_1^{s_1}(x)=\varphi(x+a)$ of the first mode with pseudospin $s_1$,
(ii) wavefunction $\varphi_2^{s_2}(x)=\varphi(x-a)$  of the second mode with pseudospin $s_1$ with  $\varphi(x)$ as in \cref{eq:gauss,eq:sech,eq:witch},
(iii) and (iv) associated field $m(x)$ and the corresponding associated field $v(x)$ for $s_1s_2=\pm1$ satisfying \cref{eq:diffeq}.
In general, for couples of modes with real wavefunction, the associated fields are not continuous (and thus not physical). 
}
   \label{fig:doublereal}
\end{figure}

\begin{figure}[tbp]
   \centering\includegraphics{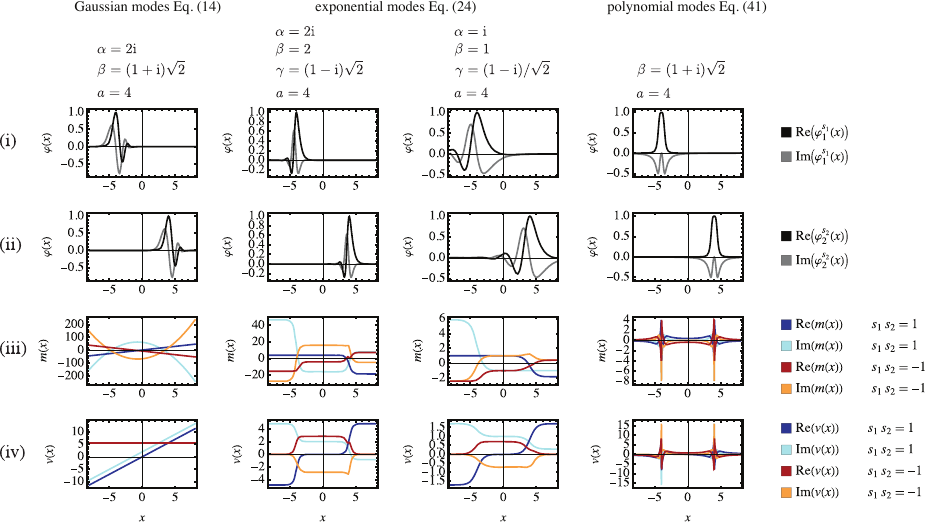}
   \caption{
Same as in \cref{fig:doublereal} but for modes with nonuniform complex phases and nonhermitian fields.
}
   \label{fig:doublecomplex}
\end{figure}

We now consider the case of two modes with identical wavefunctions centered around two distinct points at a finite distance $2a$, i.e., $\varphi(x+a)$ and $\varphi(x-a)$ with $\varphi(x)$ as in \cref{eq:gauss,eq:sech,eq:witch}, and calculated the associated fields $m(x)$ and $v(x)$ via \cref{eq:inversemu2modes,eq:inversenu2modes}.
For two Gaussian modes with same pseudospin $s$, with $\varphi(x)$ given by \cref{eq:gauss}, then the associated fields are given by
\begin{align}
m(x) &= 
-
\left(\alpha +\beta  \left(x-a\right)\right) \left(\alpha +\beta  \left(x+a\right)\right)
-\beta 
,\\
s v(x) &= \beta x + \alpha
,
\end{align}
which are polynomials of the second and first order, respectively.
On the other hand, for two Gaussian modes with opposite pseudospin $s$ and $-s$, then the fields are given by
\begin{align}
m(x) &= 
\left(\alpha +\beta  \left(x-a\right)\right) \left(\alpha +\beta  \left(x+a\right)\right)-\beta
,
\\
s v(x) &= \beta a
,
\end{align}
which are polynomials of the second and zeroth order, respectively in $x$.

For two exponentially-decaying modes with same pseudospin $s$, with $\varphi(x)$ given by \cref{eq:sech}, then the associated fields are given by
\begin{align}
m(x) &= 
- \beta^2 \gamma (\gamma + 1) y_1 y_2 
- \alpha \beta (\gamma + 1) \left(y_1 + y_2\right) 
 - \alpha^2 
 - \beta^2 \gamma 
,\\
s v(x) &= 
 \frac{1}{2} \beta (\gamma + 1) \left(y_1 + y_2\right)+\alpha 
,
\end{align}
with $y_1=\tanh(\beta (x+a))$ and $y_2=\tanh(\beta (x-a))$.
On the other hand, for two exponentially decaying modes with opposite pseudospin $s$ and $-s$, then the fields are given by
\begin{align}
m(x) &= 
\beta ^2 \gamma (\gamma +1) y_1 y_2
+\alpha  \beta  (\gamma -1) (y_1+y_2)
+\frac{\alpha ^2 (\gamma -2)}{\gamma }
-\beta ^2 \gamma
+\frac{2 \alpha  \left((\alpha +\beta  \gamma  y_1)^2+(\alpha +\beta  \gamma  y_2)^2\right)}{\gamma  (2 \alpha +\beta  \gamma  (y_1+ y_2))}
,
\\
s v(x) &= 
\frac{1}{2} \beta  (\gamma +1) (y_1-y_2)
\frac{\alpha  (\beta  \gamma  y_2-\beta  \gamma  y_1)}{\gamma  (2 \alpha +\beta  \gamma  (y_1+y_2))}
,
\end{align}
which diverge at $x\to x^*$ in the case that $\alpha\neq0$ if $2\alpha +\beta  \gamma  (y_1+ y_2)=0$ evaluated at $x=x^*$.

For two polynomially decaying modes with same pseudospin $s$, with $\varphi(x)$ given by \cref{eq:witch}, then the associated fields are given by
\begin{align}
m(x) &= 
-\frac{2 \beta ^2 \left(3 \beta ^4 \left(x^2-a^2\right)^2+2 \beta ^2 \left(a^2-3 x^2\right)-1\right)}{\left(\beta ^2 (x-a)^2+1\right) \left(\beta ^2 (x+a)^2+1\right) \left(\beta ^2 \left(x^2-a^2\right)-1\right)}
,\\
s v(x) &= 
\frac{\beta ^2 x \left(3 \beta ^4 \left(x^2-a^2\right)^2-2 \beta ^2 \left(x^2+a^2\right)-5\right)}{\left(\beta ^2 (x-a)^2+1\right) \left(\beta ^2 (x+a)^2+1\right) \left(\beta ^2 \left(x^2-a^2\right)-1\right)}.
\end{align}
On the other hand, for two polynomially decaying modes with opposite pseudospin $s$ and $-s$, then the fields are given by
\begin{align}
m(x) &= 
\frac{2 \beta ^2 \left(3 \beta ^4 \left(x^2-a^2\right)^2+2 \beta ^2 \left(x^2-3 a^2\right)-1\right)}{\left(\beta ^2 (x-a)^2+1\right) \left(\beta ^2 (x+a)^2+1\right) \left(\beta ^2 \left(x^2-a^2\right)+1\right)},
\\
s v(x) &= 
-\frac{\beta ^2 a \left(3 \beta ^4 \left(x^2-a^2\right)^2-2 \beta ^2 \left(x^2+a^2\right)-5\right)}{\left(\beta ^2 (x-a)^2+1\right) \left(\beta ^2 (x+a)^2+1\right) \left(\beta ^2 \left(x^2-a^2\right)+1\right)}.
\end{align}
In both cases of same and opposite pseudospin, the fields $m(x)$ and $v(x)$ diverge for $x\to \pm x^*$ with $x^*={\sqrt{\beta ^2 {a}^2+1}}/{\beta }$ if $x^*\in\mathbb{R}$, and are thus not physical.

The wavefunction of two identical modes and associated fields are shown in \cref{fig:doublereal,fig:doublecomplex} for the cases of real and complex wavefunctions, respectively.

\subsection{Periodic modes}

\begin{figure}[tbp]
   \centering\includegraphics{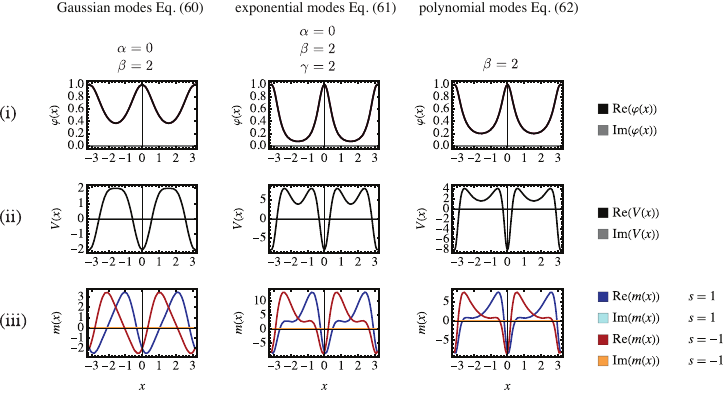}
   \caption{
Periodic modes with real wavefunction and Gaussian, exponential, or polynomial decay on the infinite interval $-\infty< x<\infty$ and associated fields:
(i) wavefunction $\varphi(x)$ as in \cref{eq:periodicgauss,eq:periodicsech,eq:periodicwitch},
(ii) associated field $V(x)$ satisfying \cref{eq:SchrodingerGeneral},
(iii) associated field $m(x)$ for $s=\pm1$ satisfying \cref{eq:diffeq} for constant field $v(x)=v=\const$ (where $v=1$).
}
   \label{fig:realperiod}
   \centering\includegraphics{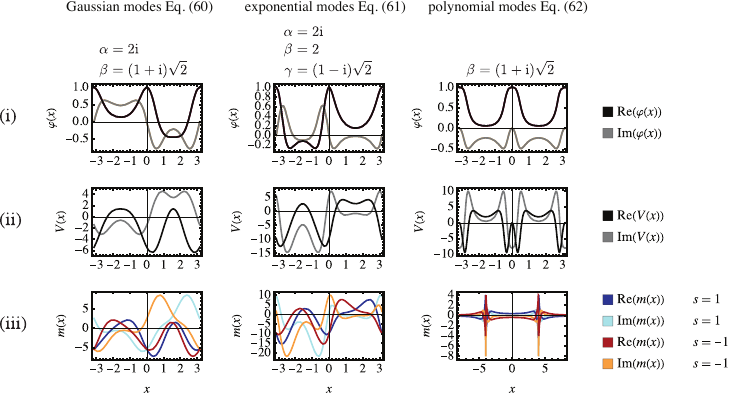}
   \caption{
Same as in \cref{fig:realperiod} but for modes with nonuniform complex phases and nonhermitian fields.
}
   \label{fig:complexperiod}
\end{figure}

We now consider periodic modes with wavefunction given by 
$\varphi(\sin{x})$,
with $\varphi(x)$ as in \cref{eq:sech,eq:gauss,eq:witch}, 
specifically,
\begin{align}
\label{eq:periodicgauss}
\varphi(x)=&
\ee^{- \frac{1}{2} \beta \sin^2(x) - \alpha \sin (x) }
,\\
\label{eq:periodicsech}
\varphi(x)=&
\ee^{ - \alpha \sin (x)} \sech^{\gamma }(\beta \sin (x))
,\\
\label{eq:periodicwitch}
\varphi(x)=&
\frac{1}{\beta^2 \sin^2{x} + 1},
\end{align}
with finite norms on the interval $-\pi\le x\le\pi$.
The wavefunction $\varphi(x)$ represents a single multilocational mode with probability density peaking on the points $x=n\pi$ with $n\in\mathbb{Z}$ which form an infinite and periodic one-dimensional array of equally-spaced points.
The probability density decays exponentially, Gaussian-like, and polynomially in the cases considered.

For periodic Gaussian modes in \cref{eq:periodicgauss}, the associated field $V(x)$ satisfying \cref{eq:SchrodingerGeneral} calculated via \cref{eq:inverseV} is given by
\begin{equation}
V(x) - E = 
\left[\alpha^2 - \beta + \beta \sin{x} (2 \alpha + \beta \sin{x})\right] \cos^2{x} 
+  (\alpha + \beta \sin{x})\sin{x}
,
\end{equation}
while the associated field $m(x)$ satisfying \cref{eq:diffeq} calculated via \cref{eq:inversemu} is given by
\begin{equation}
m(x) = V(x) - E - s [ 2 \cos{x} (\alpha + \beta \sin{x})]  v(x).
\end{equation}
For periodic and exponentially decaying modes in \cref{eq:periodicsech}, the associated field $V(x)$ is given by
\begin{equation}
V(x) - E = 
\beta^2 \gamma (\gamma + 1) y^2 \cos^2(x) + 
\left(\beta \gamma \sin (x) - 2 \alpha \beta \gamma \cos^2(x)\right) y
+ \alpha^2 \cos^2(x) - \alpha \sin (x) - \beta^2 \gamma \cos^2(x)
,
\end{equation}
while the associated field $m(x)$ is given by
\begin{equation}
m(x) = V(x) - E - 2s \left(\beta \gamma \cos (x) y -  \alpha \cos (x) \right) v(x)
,
\end{equation}
where $y=\tanh(\beta \sin{x})$.
For periodic and polynomially decaying modes in \cref{eq:periodicwitch}, the associated field $V(x)$ is given by
\begin{equation}
V(x) - E = 
-\frac{
\beta^4 \cos(4x) 
+ 2\beta ^2\left(\beta^2+2\right) \cos (2 x) 
-3\beta^4
}
{2 \left(\beta ^2 \sin ^2(x)+1\right)^2}
,
\end{equation}
while the associated field $m(x)$ is given by
\begin{equation}
m(x) = V(x) - E + s 
\frac{\beta^2 \sin (2 x) \left(\beta^2 \cos (2 x) - \beta^2 - 2\right)v(x)}
{( \beta^2 \sin^2(x) +1)^2}.
\end{equation}
All these fields are continuous and bounded.
For periodic modes, the associated field $v(x)$ satisfying \cref{eq:diffeq} obtained via \cref{eq:inversenu} for a given $m(x)$ is, in general, not continuous.

The wavefunction of periodic modes and associated fields are shown in \cref{fig:realperiod,fig:complexperiod} for the cases of real and complex wavefunctions, respectively.

Polynomially decaying modes of the Schödinger equation correspond to cases where the potential exhibits a localized dip and converges to zero for $x\to\pm\infty$, while polynomially decaying modes of the modified Jackiw-Rebbi equation correspond to cases where the mass exhibits a localized dip and converges to zero for $x\to\pm\infty$ or where the Dirac velocity diverges asymptotically.
In particular, if the Dirac velocity is uniform, then the mass converges to zero, and $\Re(m(x))$ changes signs once or three times on the real axis.
In this case, there exists only a single zero mode.
Conversely, if the mass is uniform, the Dirac velocity diverges linearly, and $\Re(v(x))$ changes sign once on the real axis.
In this case, there exists another mode with the same pseudospin $s$, which, however, decays faster than exponentially.

\section{Conclusions}

In conclusion, we employed an inverse approach to explore zero energy modes of the modified Jackiw-Rebbi and energy modes of the Schrödinger equations, uncovering several fundamental properties governing their spatial localization properties. 
In particular, we found that faster-than-exponential modes correspond to the divergence of the mass term, exponential modes correspond to finite and nonzero asymptotic masses, while slower-than-exponential modes correspond to the mass vanishing at large distances.
These results are consistent with the interpretation that the localization length is inversely proportional to the mass term. 
These three cases correspond, respectively, to the localization length converging to zero, to a finite constant value, and diverging.
We also determine the conditions where multiple zero modes localized at finite distances can exist,
and show examples of multilocational periodic modes, with probability density peaking on an array of equally-distanced points.
The different scenarios described here correspond to physical setups that are typically encountered in topological insulators, superconductors, and superfluids, e.g., in the presence of sharp of smooth topological phase boundaries or in the presence of harmonic potentials.
These properties extend naturally to nonhermitian systems.
In particular, in this context, we find that the real eigenmodes of both the modified Jackiw-Rebbi and Schrödinger equations can arise even when the associated fields (mass, Dirac velocity, or potential) are complex.
We finally note that the decaying behavior of zero modes can be experimentally determined in topological insulators and superconductors in setups where the density of states can be probed locally, e.g., by scanning tunneling microscopy (STM) in 1D topological superconductors realized with atomic chains on superconducting surfaces~\cite{pawlak_probing_2016}.
Moreover, if the edge modes have a finite electric charge, the decaying behavior can be determined by locally measuring the charge expectation value.

This work deepens the understanding of the localization properties of boundary modes localized at the phase boundary between topologically inequivalent phases.

\begin{acknowledgments} 
P.~M. is supported by the Japan Science and Technology Agency (JST) of the Ministry of Education, Culture, Sports, Science and Technology (MEXT), JST CREST Grant~No.~JPMJCR19T2, the Japan Society for the Promotion of Science (JSPS) Grant-in-Aid for Early-Career Scientists KAKENHI Grant~No.~JP23K13028 and No.~JP20K14375.
\end{acknowledgments}

\end{document}